%% file: ANA-HDBS-2019-33-PAPER.tex
\documentclass[PAPER, atlasdraft=false, texlive=2023, UKenglish, cernpreprint, texmf, orcidlogo]{atlasdoc}
 
\usepackage[subfigure]{atlaspackage}
\usepackage{atlasbiblatex}
 
\usepackage{atlasphysics}

\graphicspath{{logos/}{figures/}}
 
\newcommand{\tabscript}[3]{
\setlength{\fboxrule}{0pt}
\fbox{\ensuremath{#1^{#2}_{#3}}}
}

\AtlasTitle{AtlasTitle: Bare bones ATLAS document}
 
\AtlasVersion{3.1}

\AtlasAbstract{
Searches for the exclusive decays of the Higgs boson to
an $\omega$ meson and a photon or a $K^{*}$ meson and a photon can probe flavour-conserving and flavour-violating Higgs boson couplings to light quarks, respectively.
Searches for these decays, along with the analogous $Z$ boson decay to an $\omega$ meson and a photon,
are performed with a $pp$ collision data sample corresponding to integrated luminosities of up to
134~fb$^{-1}$ collected at $\sqrt{s}=13$~TeV with the ATLAS detector at
the CERN Large Hadron Collider.
The obtained 95$\%$ confidence-level upper limits on the respective branching fractions are
${\cal B}(H\rightarrow\omega\gamma)< 5.5\times 10^{-4}$,
${\cal B}(H\rightarrow K^{*}\gamma)< 2.2\times10^{-4}$ and
${\cal B}(Z\rightarrow \omega\gamma)<3.9\times 10^{-6}$.
The limits for $H\rightarrow \omega\gamma$ and $Z\rightarrow \omega\gamma$ are 370 times and 140 times the Standard Model expected values, respectively.
The result for $Z\rightarrow \omega\gamma$ corresponds to a two-orders-of-magnitude improvement over the limit obtained by the DELPHI experiment at LEP. }
 
\AtlasRefCode{HDBS-2019-33}
 
\PreprintIdNumber{CERN-EP-2022-288}

\AtlasJournal{Phys.\ Lett.\ B.}
\AtlasJournalRef{Phys.\ Lett.\ B. 847 (2023) 138292}
\AtlasDOI{10.1016/j.physletb.2023.138292}

\hypersetup{pdftitle={ATLAS document},pdfauthor={The ATLAS Collaboration}}
 
\begin{document}
 
\AtlasTitle{Search for exclusive Higgs and $Z$ boson decays to $\omega\gamma$ and Higgs boson decays to $K^{*}\gamma$ with the ATLAS detector}
 
\maketitle

\section{Introduction}
\label{sec:intro}
 
The discovery of a Higgs boson $H$ with a mass of approximately $m_{H}
= 125$~\GeV~\cite{HIGG-2012-27,CMS-HIG-12-028} by the ATLAS and CMS collaborations at
the Large Hadron Collider (LHC) has led to a variety of measurements
to determine its properties, which so far have found no deviations
from the Standard Model
(SM)~\cite{HIGG-2021-23,CMS-HIG-22-001}. In the SM,
fermion masses are generated via Yukawa interactions between the Higgs
and fermion fields. The Higgs boson couplings to the fermions are subject
to substantial modifications in
various theories beyond the SM~\cite{Heinemeyer:2013tqa}.
Theories that modify the couplings of the Higgs boson to quarks
include the Minimal Flavour Violation
framework~\cite{DAmbrosio:2002ex}, the Froggatt--Nielsen
mechanism~\cite{Froggatt:1978nt}, the Higgs-dependent Yukawa couplings
model~\cite{Giudice:2008uua}, the Randall--Sundrum family of
models~\cite{Randall:1999ee}, and the possibility of the Higgs boson
being a composite pseudo-Goldstone boson~\cite{Dugan:1984hq}.
Measurements of Higgs boson production in association with top-quark
pairs~\cite{HIGG-2015-07} and decays of Higgs bosons into pairs of
$\tau$-leptons~\cite{HIGG-2015-07,HIGG-2013-32,CMS-HIG-13-004} or
bottom-quarks~\cite{HIGG-2016-29,CMS-HIG-16-044} have been performed
and were found to be consistent with the SM expectations, providing clarity on third-generation fermion
interactions with the Higgs boson.
Evidence has been reported for the Higgs boson coupling to the second
generation through the $H\to\mu^{+}\mu^{-}$ decay~\cite{CMS-HIG-19-006,HIGG-2019-14},
but there is no further experimental evidence of interactions with fermions of the first and second generations.
Searches have also been performed for the decay $H\to c\bar{c}$ by the ATLAS and CMS collaborations \cite{ATLAS:2018mgv,ATLAS:2022ers,CMS:2019hve,CMS-HIG-21-008}.
 
\begin{figure}[!htbp]
\centering
\subfigure[]{\includegraphics[width=0.35\textwidth]{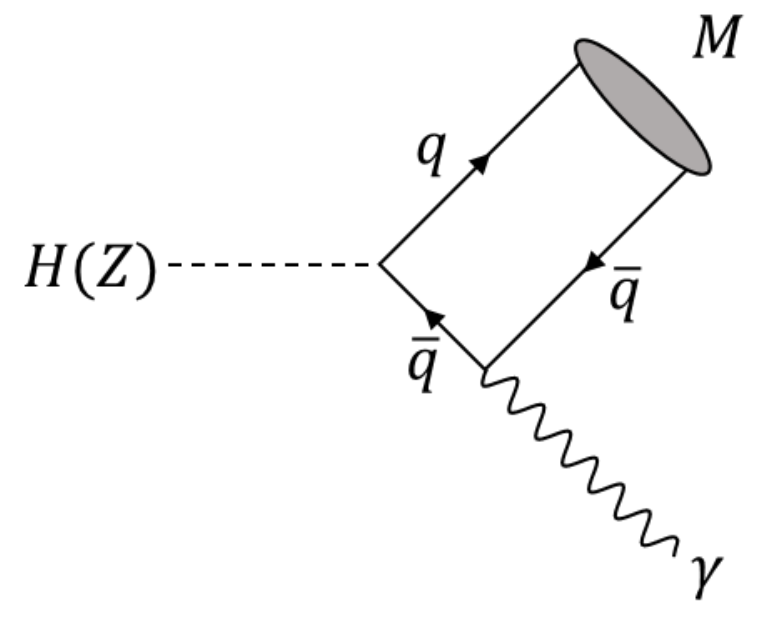}\label{fig:feynman2}}
\subfigure[]{\includegraphics[width=0.35\textwidth]{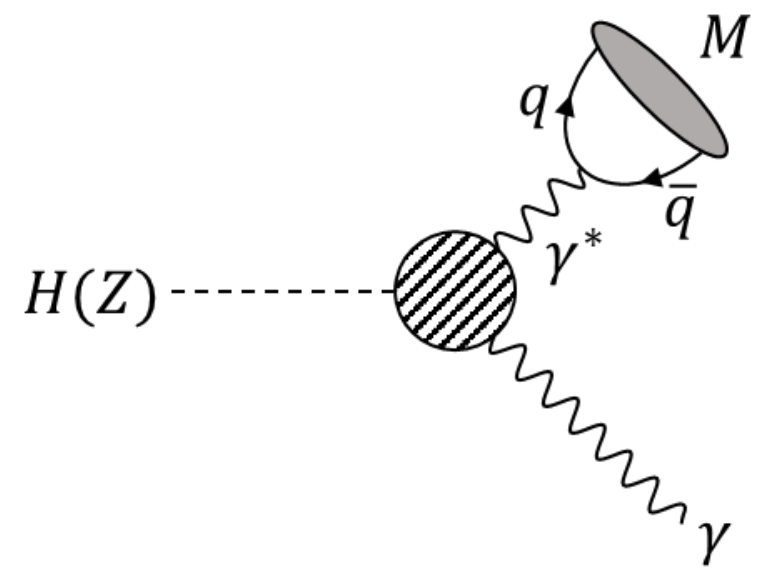}\label{fig:feynman3}}
\caption{Feynman diagrams for the~\subref{fig:feynman2} direct and~\subref{fig:feynman3} indirect contributions to the $H/Z\to\mathcal{M}\gamma$ decays.}
\label{fig:Feynman}
\end{figure}

Currently, the light ($u$, $d$, $s$) quark couplings to the Higgs
boson are loosely constrained by existing data on the total Higgs
boson width and combined measurements of Higgs boson production and
decays~\cite{HIGG-2021-23,CMS-HIG-22-001}, while the large multi-jet background at
the LHC inhibits the study of such couplings with inclusive $H\to
q\bar{q}$ decays.
Rare decays of the Higgs boson into a meson and a photon have been suggested as a probe
of the couplings of the Higgs boson to light quarks.
Higgs boson decays into a heavy quarkonium state, \Jpsi, \psip and \Ups{$n$} with $n=1,2,3$, and a photon have
been suggested for probing the charm- and bottom-quark couplings to
the Higgs boson~\cite{VatoHBranching,HBranching,Bodwin:2014bpa,Grossmann:2015lea}. These
have already been searched for by the ATLAS and CMS
collaborations~\cite{HIGG-2014-03,HIGG-2016-23,HDBS-2018-53-new,CMS-HIG-14-003,CMS-SMP-17-012}.  Higgs boson decays into a $\phi$ or $\rho$ meson and a
photon are potential probes for the $u$, $d$ and $s$ quark
couplings to the Higgs boson~\cite{Koenig:2015pha,Kagan:2014ila,Perez:2015lra}. Searches
have already been peformed for these decays by the ATLAS
Collaboration~\cite{HIGG-2016-05,HIGG-2016-13} and the CMS Collaboration~\cite{CMS-HIG-19-012}.
The current constraints with the corresponding predictions are given in Ref.~\cite{ATL-PHYS-PUB-2023-004}.
The partial widths for these decays are driven by two (interfering)
contributions: one known as \enquote{direct}, which scales with the Yukawa
couplings, and the other known as \enquote{indirect}, which mimics $H\to\gamma\gamma$ but with
one photon fragmenting into a quark--antiquark pair, forming the meson. These contributions
are shown in Figure~\ref{fig:Feynman}.
The interference between these two amplitudes is rather strong and
typically destructive. In the case of the $\mit\Upsilon$ channel, almost complete
destructive interference occurs, reversing the expected scaling with quark mass.
 
This paper describes a search for Higgs boson decays into the
exclusive final states $\omega\gamma$ and $K^{*}\gamma$. These decays can probe
the flavour-conserving coupling of the Higgs boson to up and down quarks, and
the flavour-violating coupling of the Higgs boson to down and strange quarks, respectively.
Two calculations of the expected SM
branching fractions for the $\omega$ channel are available:
$\mathcal{B}\left(H\to\omega\,\gamma\right)=(1.48 \pm
0.08)\times10^{-6}$~\cite{Koenig:2015pha} and
$\mathcal{B}\left(H\to\omega\,\gamma\right)=(1.6\pm
0.2)\times10^{-6}$~\cite{Kagan:2014ila}. For the $K^{*}\gamma$ decay, which arises from loop contributions,
only the expected branching fraction of $H\to d\bar{s}\, + \,\bar{d}s$ is available,
with a value of $\mathcal{B}(H\to d\bar{s}\, + \,\bar{d}s)=1.19 \times
10^{-11}$~\cite{Aranda:2020tqw}.
The branching fraction for $H\to K^*\gamma$ is expected to be much smaller.
This is the first exclusive decay
analysis to target flavour-changing interactions of the Higgs boson. Earlier searches
for flavour-changing neutral currents via the top-quark decays $t\to uH$ and
$t\to cH$ were performed by the ATLAS and CMS
collaborations~\cite{HIGG-2016-26,TOPQ-2017-15,CMS-TOP-19-002,CMS-TOP-20-007-PRL},
with no evidence found for these decays.
 
The decay
$\omega\to\pi^{+}\pi^{-}\pi^{0}$ is used to reconstruct the $\omega$
meson, and the decay $K^{*}\to K^{+}\pi^{-}$ is used to reconstruct
the $K^{*}$ meson. The branching fractions of the respective meson
decays are accounted for when calculating the expected signal yields.
A search for the analogous decay of the $Z$ boson into a $\omega$
meson and a photon is also presented. The channel has been
studied theoretically~\cite{Huang:2014cxa,Grossmann:2015lea} as a
unique precision test of the SM and the factorisation approach in
quantum chromodynamics (QCD), in an environment where the power
corrections in terms of the QCD energy scale divided by the mass of the
vector boson are small~\cite{Grossmann:2015lea}. The large $Z$ boson
production cross section at the LHC means that rare $Z$ boson decays
can be probed at branching fractions much smaller than for Higgs boson
decays into the same final states. The SM branching fraction
prediction for the decay considered in this paper is
$\mathcal{B}\left(Z\to \omega\gamma\right)=(2.82\pm
0.40)\times10^{-8}$~\cite{Grossmann:2015lea}. A previous search was
performed at the DELPHI experiment, yielding an upper limit on the branching
fraction of $\mathcal{B}\left(Z\to \omega\gamma\right)< 6.5 \times 10^{-4}$ \cite{DELPHI:1994mra}.
 
\section{ATLAS detector}
\label{sec:detector}
 
The ATLAS detector~\cite{PERF-2007-01} is a multi-purpose particle
physics detector with an approximately forward-backward symmetric cylindrical
geometry and near $4\pi$ coverage in solid angle.\footnote{ATLAS uses
a right-handed coordinate system with its origin at the nominal
interaction point (IP) in the centre of the detector and the $z$-axis
along the beam pipe. The $x$-axis points from the IP to the centre of
the LHC ring, and the $y$-axis points upward. Cylindrical coordinates
$(r, \phi)$ are used in the transverse plane, $\phi$ being the
azimuthal angle around the $z$-axis. The pseudorapidity is defined in
terms of the polar angle $\theta$ as $\eta = -
\ln{\tan(\theta/2)}$.} It consists of an inner tracking
detector surrounded by a thin superconducting solenoid,
electromagnetic and hadronic calorimeters, and a muon spectrometer.
The inner tracking detector (ID) covers the pseudorapidity range
$|\eta| < 2.5$, and is surrounded by a thin superconducting solenoid
providing a 2~T magnetic field. At small radii, a high-granularity
silicon pixel detector covers the vertex region and typically provides
three measurements per track. A new innermost pixel-detector layer,
the insertable B-layer, was added before 13~\TeV\ data-taking began in
2015 and provides an additional measurement at a radius of about 33~mm
around a new and thinner beam pipe~\cite{ATLAS-TDR-19,PIX-2018-001}. The pixel
detectors are followed by a silicon microstrip tracker, which
typically provides four space-point measurements per track. The
silicon detectors are complemented by a gas-filled straw-tube
transition radiation tracker, which enables radially extended track
reconstruction up to $|\eta| = 2.0$, with typically 35 measurements
per track.
The calorimeter system covers the pseudorapidity range $|\eta| <
4.9$. A high-granularity lead/liquid-argon (LAr) sampling
electromagnetic calorimeter covers the region $|\eta| < 3.2$, with an
additional thin LAr presampler covering $|\eta| < 1.8$ to correct for
energy losses upstream. The electromagnetic calorimeter is divided
into a barrel section covering $|\eta| < 1.475$ and two endcap
sections covering $1.375 < |\eta| < 3.2$. For $|\eta| < 2.5$ it is
divided into three layers in depth, which are finely segmented in
$\eta$ and $\phi$. A steel/scintillator-tile calorimeter provides
hadronic calorimetry in the range $|\eta| < 1.7$, while in the endcap
region, $1.5 < |\eta| < 3.2$, a copper/LAr calorimeter is used. The
solid-angle coverage is completed with forward copper/LAr and
tungsten/LAr calorimeter modules in $3.1 < |\eta| < 4.9$, optimised
for electromagnetic and hadronic measurements, respectively.
The muon spectrometer surrounds the calorimeters and comprises
separate trigger and high-precision tracking chambers measuring the
deflection of muons in a magnetic field provided by three air-core
superconducting toroids.
 
A two-level trigger and data acquisition system is used to provide an
online selection and record events for offline
analysis~\cite{TRIG-2016-01}. The level-1 trigger is implemented in
hardware and uses a subset of detector information to reduce the event
rate to 100~kHz or less from the maximum LHC collision rate of 40~MHz.
It is followed by a software-based high-level trigger which
filters events using the full detector information and records events
for detailed offline analysis at an average rate of 1~kHz.
An extensive software suite~\cite{ATL-SOFT-PUB-2021-001} is used in
data simulation, in the reconstruction and analysis of real and
simulated data, in detector operations, and in the trigger and data
acquisition systems of the experiment.

\section{Data and Monte Carlo simulation}
\label{sec:DataMC}
 
The searches are performed in two distinct decay modes, $\omega\gamma$
and $K^{*}\gamma$, using a $pp$ collision data sample
collected at a centre-of-mass energy $\sqrt{s} = 13$~\TeV. Following
the requirement that events must be collected under stable LHC beam
conditions and that all relevant detector components are in good
operating condition, the total integrated luminosity available is 89.5
and 134~fb$^{-1}$ for the $\omega\gamma$ and $K^{*}\gamma$ final
states, respectively, with an uncertainty of 1.7$\%$ \cite{DAPR-2013-01, ATLAS-CONF-2019-021}.
The data samples were recorded by a combination of dedicated triggers,
which were integrated into the trigger menu at different times.
This is reflected in the different integrated luminosities.
Namely, the trigger for the $\omega\gamma$ final state recorded data from July 2017 to the end of data taking in 2018, whilst the triggers for the $K^{*}\gamma$ final state recorded data from May 2016 to the end of data taking in 2018.
These require a photon at the level-1 trigger, and a photon and an isolated pair of ID
tracks at the high-level trigger.
At the high-level trigger, an isolated photon with a transverse
momentum $\pt^{\gamma} > 35$~\GeV~\cite{TRIG-2016-01} is required in general, with the exception of data recorded for the $K^{*}\gamma$ final state in 2017--2018, where the threshold was reduced to $\pt^{\gamma} > 25$~\GeV.
For the ID tracks a modified version of the
$\tau$-lepton trigger algorithms~\cite{ATLAS-CONF-2017-061} is used.
The triggers for the $K^{*}\gamma$ final state require at least one track to have a $\pt$ greater than 15~\GeV,
whilst the triggers for the $\omega\gamma$ final state require at least one track to have a $\pt$ greater than 25~\GeV.
In each case, the track is required to be associated with a topological cluster of calorimeter cells~\cite{PERF-2014-07}
with a transverse energy greater than 25~\GeV.
Different requirements on the invariant mass of the pair of tracks are applied, depending on
the mass of each targeted meson. For
$\omega\to\pi^{+}\pi^{-}\pi^{0}$
an invariant mass of the pair of tracks in the range 279 to 648~\MeV\ is required,
under the charged-pion hypothesis.
For $K^{*}\to K^{+}\pi^{-}$ an invariant mass of 790 to 990~\MeV, under the
$K^{\pm}\pi^{\mp}$ hypothesis, is required.
The trigger efficiency with respect to the offline selection, as described in Section~\ref{sec:selection},
is approximately $78\%$ for the $K^*\gamma$ final state and $52\%$ for the $\omega\gamma$ final state.
 
Higgs boson production through the gluon–gluon fusion ($ggH$) and vector-boson fusion (VBF) processes was modelled up to next-to-leading order (NLO) in $\alphas$ using the \POWHEGBOX[v4] Monte Carlo (MC) event generator~\cite{powheg1,powheg2,powheg3,powheg4,powheg5}. \POWHEGBOX[v4] was interfaced with the \Pythia[8.244] MC event generator~\cite{Sjostrand:2007gs,Sjostrand:2006za} to model the parton shower, hadronisation and underlying event, with parameter values set according to the AZNLO tune~\cite{STDM-2012-23} and using CTEQ6L1 parton distribution functions (PDFs)~\cite{Pumplin:2002vw}. Additional contributions from the associated production of a Higgs boson and a $W^{\pm}$ or $Z$ boson (denoted by $WH$ and $ZH$, respectively) were modelled by \Pythia[8.244]  with \NNPDF[2.3lo] PDFs~\cite{Ball:2012cx} and the A14 tune for hadronisation and the underlying event~\cite{ATL-PHYS-PUB-2014-021}. Higgs boson production through associated production with top quarks ($t\bar{t}H$) was modelled using \Pythia[8.244], and \AMCatNLO to model the parton shower~\cite{Alwall:2014hca}, again with the \NNPDF[2.3lo] PDFs and A14 tune. The production rates and kinematic distributions for the SM Higgs boson with $m_H = 125$~\GeV\ are assumed throughout. These were obtained from Ref.~\cite{LHCHCWG:2012vm} and are summarised below. The $ggH$ production rate is normalised such that it reproduces the total cross section predicted by a next-to-next-to-next-to-leading-order QCD calculation with NLO electroweak corrections applied~\cite{Actis:2008ug,Anastasiou:2008tj}. The VBF production rate is normalised to an approximate next-to-next-to-leading-order (NNLO) QCD cross section with NLO electroweak corrections applied~\cite{Ciccolini:2007jr,Ciccolini:2007ec,Bolzoni:2010xr}. The $WH$ and $ZH$ production rates are normalised to cross sections calculated at NNLO in QCD with NLO electroweak corrections~\cite{Brein:2003wg} including the NLO QCD
corrections for $gg\to ZH$.
\POWHEGBOX[v4] was also used to
model inclusive $Z$ boson production with CT10 PDFs. \Pythia[8.244] with CTEQ6L1 PDFs and
the AZNLO tune was used to simulate the parton showering and hadronisation. The prediction is
normalised to the total cross section obtained from the measurement in Ref.~\cite{STDM-2015-03}. The Higgs and $Z$ boson decays were simulated as a cascade of two- and three-body decays, accounting for angular momentum conservation. The meson invariant mass distributions were simulated by \Pythia. The branching
fraction for the decay $\omega\to\pi^{+}\pi^{-}\pi^{0}$ is $(89.2 \pm 0.7)\%$.
The decay $K^{*}\to K\pi$ has a branching fraction close to 100$\%$, of which two-thirds correspond to decays to a charged kaon and a charged pion~\cite{PDG}.
The simulated events were passed through a detailed \GEANT simulation of the ATLAS detector~\cite{GEANT4,2010wqa} and processed with the same software used to reconstruct the data. Simulated additional $pp$ collisions in the same or neighbouring bunch crossings (pile-up events) are also included and the distribution of these is matched to the conditions observed in the data.
 
\section{Object and event selections}
\label{sec:selection}
 
The similarities of the final states in these searches allow common event selections to be used. A pair of oppositely charged reconstructed ID tracks is required for both, with the addition of a neutral pion in the reconstruction of the $\omega\gamma$ final state.
 
Events with a $pp$ interaction vertex reconstructed from at least two ID tracks with $\pt > 500$~\MeV\ are considered in the analysis. Within an event, the primary vertex is defined as the reconstructed vertex with the largest $\sum{\pt^{2}}$ of associated ID tracks.
 
Photons are reconstructed from clusters of energy in the electromagnetic calorimeter. Reconstructed photon candidates are required to have $\pt^{\gamma} > 35$~\GeV, $|\eta^{\gamma}| < 2.37$,
excluding the barrel/endcap calorimeter transition region $1.37 < |\eta^{\gamma}| < 1.52$, and to satisfy \enquote{tight} photon
identification criteria~\cite{PERF-2017-02}. An isolation requirement is imposed to further suppress contamination from
jets. The sum of the transverse momenta of all tracks within $\Delta R = \sqrt{(\Delta\phi)^2 + (\Delta\eta)^2} = 0.2$ of the photon direction, excluding those associated with the reconstructed photon, is required to be less than 5$\%$ of $\pt^{\gamma}$.
Moreover, the sum of the transverse momenta of all calorimeter energy deposits within $\Delta R = 0.4$ of the photon direction, excluding those associated with the reconstructed photon, is required to be less than $2.45~\GeV + 0.022 \times \pt^{\gamma}$.
To mitigate the effects of multiple $pp$ interactions in the same or neighbouring bunch crossings, only ID tracks consistent with originating from the primary vertex are considered in the photon track-based isolation. For the calorimeter-based isolation the effects of the underlying event and multiple $pp$ interactions are also accounted for on an event-by-event basis using an average underlying-event energy density determined from data~\cite{PERF-2017-02}.
 
Charged-hadron candidates are reconstructed from ID tracks which are required to have $|\eta| < 2.5$, $\pt > 3$~\GeV\ and to satisfy basic quality criteria, including a requirement on the number of hits in the silicon detectors~\cite{ATL-PHYS-PUB-2015-051}. For the $\omega\gamma$ analysis, the combination of a pair of charged-hadron candidates with opposite charge and a neutral-pion candidate is denoted by $\mathcal{M}$, whereas for the $K^{*}\gamma$ analysis, $\mathcal{M}$ denotes the combination of a pair of charged-hadron candidates with opposite charge. Within a pair, the charged-hadron candidate with the higher $\pt$, referred to as the leading charged-hadron candidate, is required to have $\pt > 20$~\GeV for the $K^{*}\gamma$ analysis and $\pt > 25$~\GeV for the $\omega\gamma$ analysis. The selection of candidates for the charged-particle pair in $\mathcal{M}$ is based on their invariant masses. Charged-hadron pairs satisfying these requirements are assumed to be a $\pi^+\pi^-$ pair in the $\omega\gamma$ analysis. In the $K^{*}\gamma$ channel, tracks are assigned a particular mass hypothesis by calculating the invariant mass of the di-track system for both possible assignments ($K/\pi$ or $\pi/K$). The combination which results in an invariant mass closer to $m_{K^{*}}$ is chosen. Additionally, the $\pt$ of the di-track system must satisfy $\pt > 35$~\GeV. For the $\omega\gamma$ channel, charged-hadron pairs with an invariant mass of 279 to 648~\MeV\ are selected as candidates. For the $K^{*}\gamma$ channel, charged-hadron pairs with an invariant mass of 790 to 990~\MeV\ are selected as candidates. Selected charged-hadron pair candidates are required to satisfy an isolation requirement: the sum of the $\pt$ of the reconstructed ID tracks from the primary vertex that are within $\Delta R = 0.2$ of the leading charged-hadron candidate (excluding from the sum the charged-hadron candidate defining the pair) is required to be less than 10$\%$ of the $\pt$ of the charged-hadron pair candidate for the $\omega\gamma$ channel, and less than 20$\%$ for the $K^{*}\gamma$ channel.
 
Neutral-pion candidates are expected to leave a signature of a pair of overlapping clusters in the electromagnetic calorimeter. In the case of $H\to\omega\gamma$ decays, the $\pi^{0}$ is expected to be very close in $\Delta R$ to the $\pi\pi$ system. Tau particle-flow object (TauPFO) algorithms~\cite{PERF-2014-06} optimised to search for neutral pions in $\tau$-like decay signatures, using boosted decision trees, are used to reconstruct the neutral pion in the $\omega$ decay. The $\tau$-jets are searched for, with no identification requirements, within $\Delta R = 0.1$ of the $\pi^{+}\pi^{-}$ system. The presence of neutral-pion particle-flow objects is then examined within each $\tau$-jet object. The closest particle-flow object consistent with a $\pi^{0}$ is taken to be the $\pi^{0}$ candidate, which is then added to the charged-hadron candidates to form $\mathcal{M}$ for the $\omega\gamma$ channel. An additional mass requirement of 650 to 850~\MeV\ is then imposed on the fully reconstructed $\omega$ meson.
 
The $\mathcal{M}$ candidates are subsequently combined with the reconstructed photons. When multiple combinations are possible, a situation that arises only in a few percent of the events, the combination of the highest-$\pt$ photon and the ${\cal M}$ candidate with an invariant mass closest to the respective meson mass is selected. The event is retained for further analysis if the requirement  $\Delta\phi(\mathcal{M},\gamma) > \pi/2$ on the azimuthal angular separation between the meson candidate and the photon is satisfied. This selection defines the \enquote{signal region}.
 
For the $\omega\gamma \to \pi^{+}\pi^{-}\pi^{0}\gamma$ final state, the total signal efficiency (kinematic acceptance, trigger and reconstruction efficiencies) is 2.2$\%$ and 0.4$\%$ for the Higgs and $Z$ boson decays, respectively. The corresponding
efficiency for the $K^{*}\gamma$ final state in Higgs boson decay is 12.1$\%$.\footnote{This efficiency is estimated with respect to $K^*$ decays to a charged kaon and a charged pion.} The difference in efficiency between the Higgs
and $Z$ boson decays arises primarily from the softer $\pt$ distributions of the photon and charged-hadron
candidates from $Z\to\omega\gamma$ production.
The efficiency difference between the two final states is due to the presence of the neutral pion in the decay of the $\omega$.
This arises both from the reconstruction efficiency of the neutral pion itself,
and from the softer $\pt$ distributions of the charged hadrons in the decay of the $\omega$ compared to the $K^*$, caused by the neutral pion taking a portion of the decay energy.
 
The Higgs boson signal $m_{\mathcal{M}\gamma}$ distribution for the $H\to \omega\gamma$ decay is modelled with a sum of a Crystal Ball distribution and a Gaussian distribution, while for $H\to K^{*}\gamma$ a sum of two Gaussian probability density functions (pdf) with a common mean value is used. For the $Z$ boson signal the $m_{\mathcal{M}\gamma}$ distribution is modelled with a double Voigtian pdf (a convolution of relativistic Breit--Wigner and Gaussian pdfs) corrected with a mass-dependent efficiency factor.
 
Additionally, sideband meson mass datasets are defined for both analyses, where the full set of selection criteria is imposed except for different invariant mass requirements on $\mathcal{M}$. For the $H\to \omega\gamma$ decay channel, the requirement is $m_{\omega} < 650$~\MeV\ or $850~\MeV < m_{\omega}< 2000$~\MeV. For $H\to K^{*}\gamma$ the requirement is $m_{K^{*}}< 790$~\MeV\ or $m_{K^{*}} > 990$~\MeV. These requirements produce datasets that are orthogonal to the signal region, but still provide a useful description of the background processes.
 
\section{Background estimation}
\label{sec:background}
For both the $\omega\gamma$ and $K^{*}\gamma$ final states, the main sources of background in the searches are events involving inclusive $\gamma$\;\!+\,jet or multi-jet processes, where an $\mathcal{M}$ system is reconstructed from ID tracks originating from a jet. From the selection criteria discussed earlier, the shape of this background exhibits a kinematic turn-on structure in the $m_{\mathcal{M}\gamma}$ distribution around 100~\GeV, in the region of the $Z$ boson signal, and a smoothly falling background in the region of the Higgs boson signal. These processes are modelled inclusively with a non-parametric data-driven approach using templates to describe the relevant distributions~\cite{Chisholm:2021pdn}. The background normalisation and shape are simultaneously extracted from a fit to the data. A similar procedure was used in the searches for Higgs and $Z$ boson decays into $\phi\gamma$ and $\rho\gamma$~\cite{HIGG-2016-05,HIGG-2016-13} and the searches for Higgs and $Z$ boson decays into $\Jpsi\gamma$, $\psip\gamma$ and $\Ups{$n$}\gamma$~\cite{HIGG-2014-03,HIGG-2016-23,HDBS-2018-53-new}.

\subsection{Background modelling}
 
\begin{figure}[!htbp]
\centering
\includegraphics[width=0.9\textwidth]{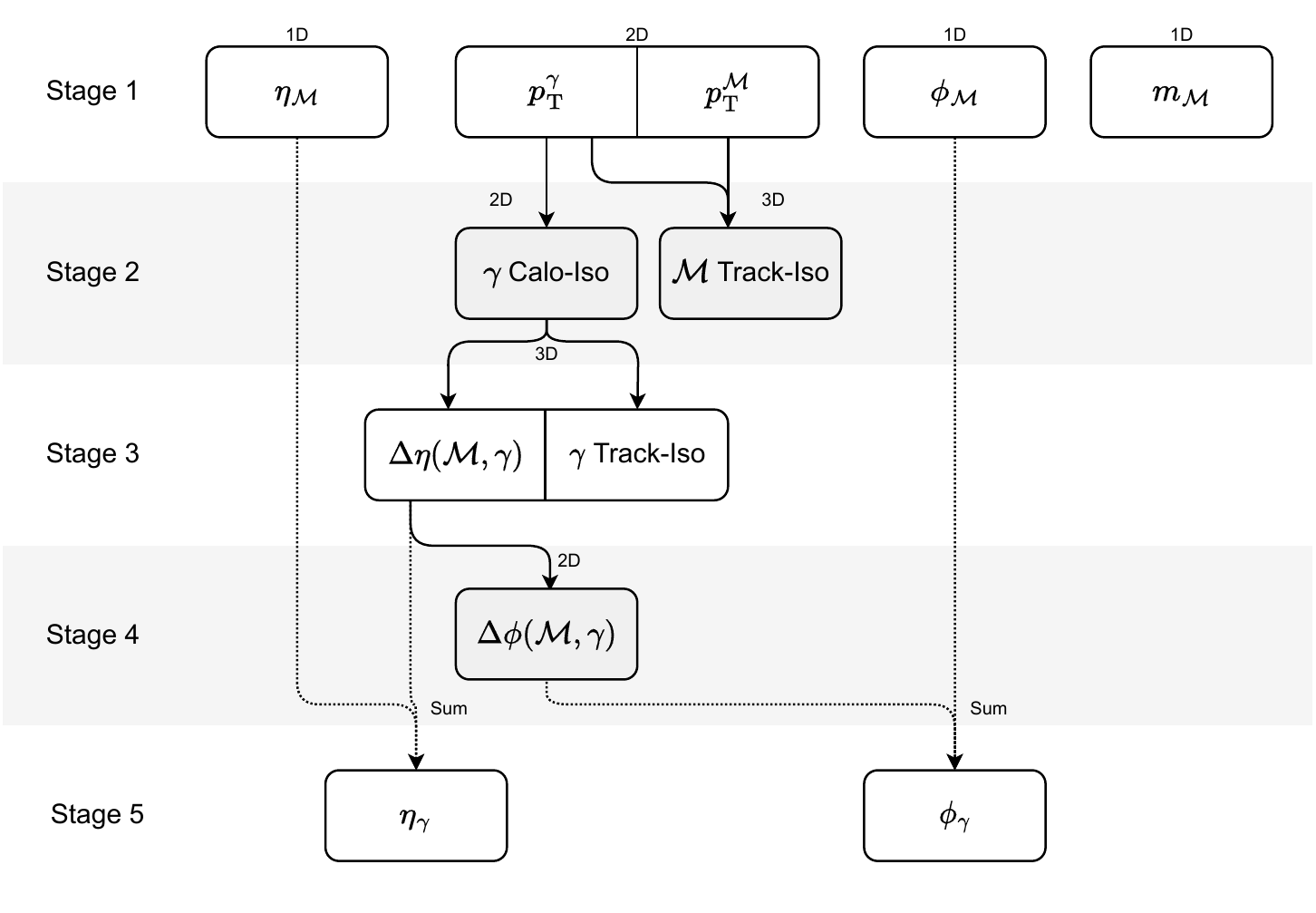}
\caption{A schematic diagram showing the stages of the background model used for the analyses presented. Variables are drawn using a combination of 1D, 2D and 3D histograms in order to retain the correlations between them. Solid arrows correspond to where variables have been sampled from the previous stage. Dashed arrows correspond to where previous variables have been summed to produce a new variable. Joined cells indicate variables that have been simultaneously sampled from the previous stage.}
\label{fig:BkgdModelSchematic}
\end{figure}

The background modelling procedure for each final state uses a sample of approximately 16\,000 $\pi^{+}\pi^{-}\pi^{0}\gamma$ and 280\,000 $K^{+}\pi^{-}\gamma$ candidate events in data. These events pass all the kinematic selection requirements described previously, except that the photon and $\mathcal{M}$  candidates are not required to satisfy the nominal isolation requirements. This selection defines the \enquote{generation region} (GR), which is background-dominated. From these events, pdfs are constructed to describe the distributions of the relevant kinematic and isolation variables and their most important correlations. In this way, in the absence of appropriate simulations, pseudocandidates are generated, from which the background shape in the discriminating variable is derived.
This ensemble of pseudocandidates is produced by randomly sampling the distributions of the relevant kinematic and isolation variables, which are estimated from the data in the GR. Each pseudocandidate is described by $\mathcal{M}$ and $\gamma$ four-momentum vectors and the associated $\mathcal{M}$ and photon isolation variables. The $\mathcal{M}$ four-momentum vector is constructed from sampled $\eta_{\mathcal{M}}$, $\phi_{\mathcal{M}}$, $m_{\mathcal{M}}$ and $\pt^{\mathcal{M}}$ values. For the $\gamma$ four-momentum vector, the $\eta_{\gamma}$ and $\phi_{\gamma}$ values are determined from the sampled $\Delta\phi(\mathcal{M},\gamma)$ and $\Delta\eta(\mathcal{M},\gamma)$ values, whereas $\pt^{\gamma}$ is sampled directly.
The most important correlations among these kinematic and isolation variables in background events are retained in the generation of the pseudocandidates.
This is achieved through the following sampling scheme, also depicted in Figure~\ref{fig:BkgdModelSchematic}, where the steps are performed sequentially, following the magnitude of the observed correlations:
 
i) Initially, values for $\pt^{\mathcal{M}}$ and $\pt^{\gamma}$ are drawn from a two-dimensional pdf of ($\pt^{\mathcal{M}}$, $\pt^{\gamma}$). The values of $\eta_{\mathcal{M}}$ and the $\phi$ angle of the meson are sampled from one-dimensional pdfs of the variables. Finally, a value for the meson mass is sampled from a one-dimensional pdf of $m_{\mathcal{M}}$.
 
ii) The isolation of the meson candidate is sampled from a three-dimensional pdf based on the values of $\pt^{\mathcal{M}}$ and $\pt^{\gamma}$ obtained in the previous step. Then, the photon's relative calorimeter isolation is sampled from a two-dimensional pdf, based on the value of $\pt^{\gamma}$ obtained in the previous step.
 
iii) From this value of the photon's relative calorimeter isolation, the values of the pseudorapidity difference between the $\mathcal{M}$ and $\gamma$ candidates, $\Delta\eta(\mathcal{M},\gamma)$, and the photon's relative track isolation are sampled simultaneously from a three-dimensional pdf.
 
iv) The value of the azimuthal angular separation, $\Delta\phi(\mathcal{M},\gamma)$, is sampled from a two-dimensional pdf based on the value of $\Delta\eta(\mathcal{M},\gamma)$ drawn in the previous step.
 
v) Given these sampled values, and the values sampled previously for $\Delta\eta(\mathcal{M},\gamma)$ and $\Delta\phi(\mathcal{M},\gamma)$, the values of $\eta_{\mathcal{M}}$ and $\phi_{\gamma}$ are then defined.
 
The nominal selection requirements are imposed on the ensemble, and the surviving pseudocandidates are used to construct templates for the $m_{\mathcal{M}\gamma}$ distribution, which are then smoothed using Gaussian kernel density estimation~\cite{Cranmer:2000du}. Signal injection tests were performed to ensure that the background model is not affected by any potential signal contamination.

\subsection{Background validation} \label{sec:BkgdValidation}
 
\begin{figure}[!htbp]
\centering
\subfigure[]{\includegraphics[width=0.35\textwidth]{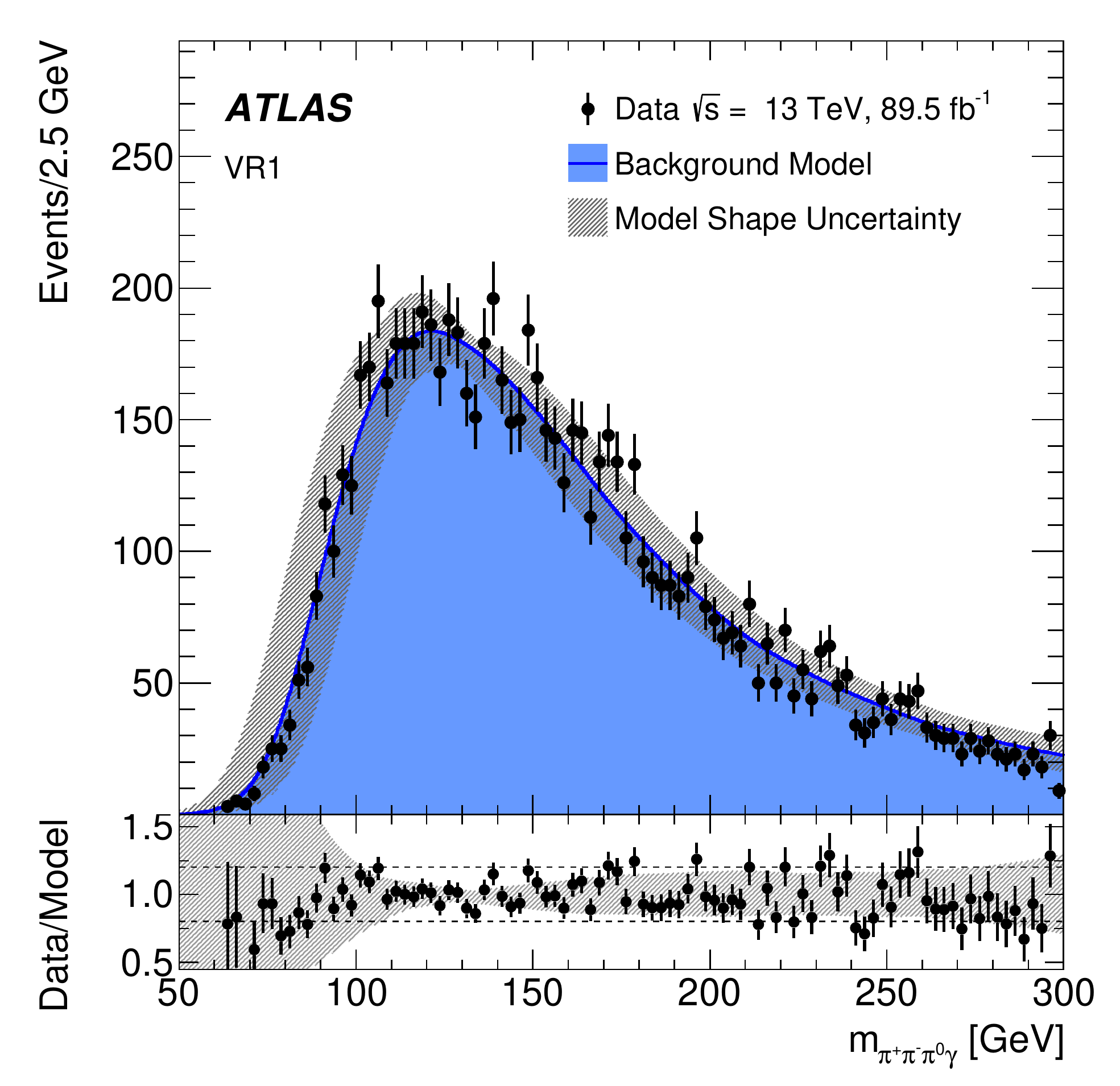}\label{fig:Omegavr1}}
\subfigure[]{\includegraphics[width=0.35\textwidth]{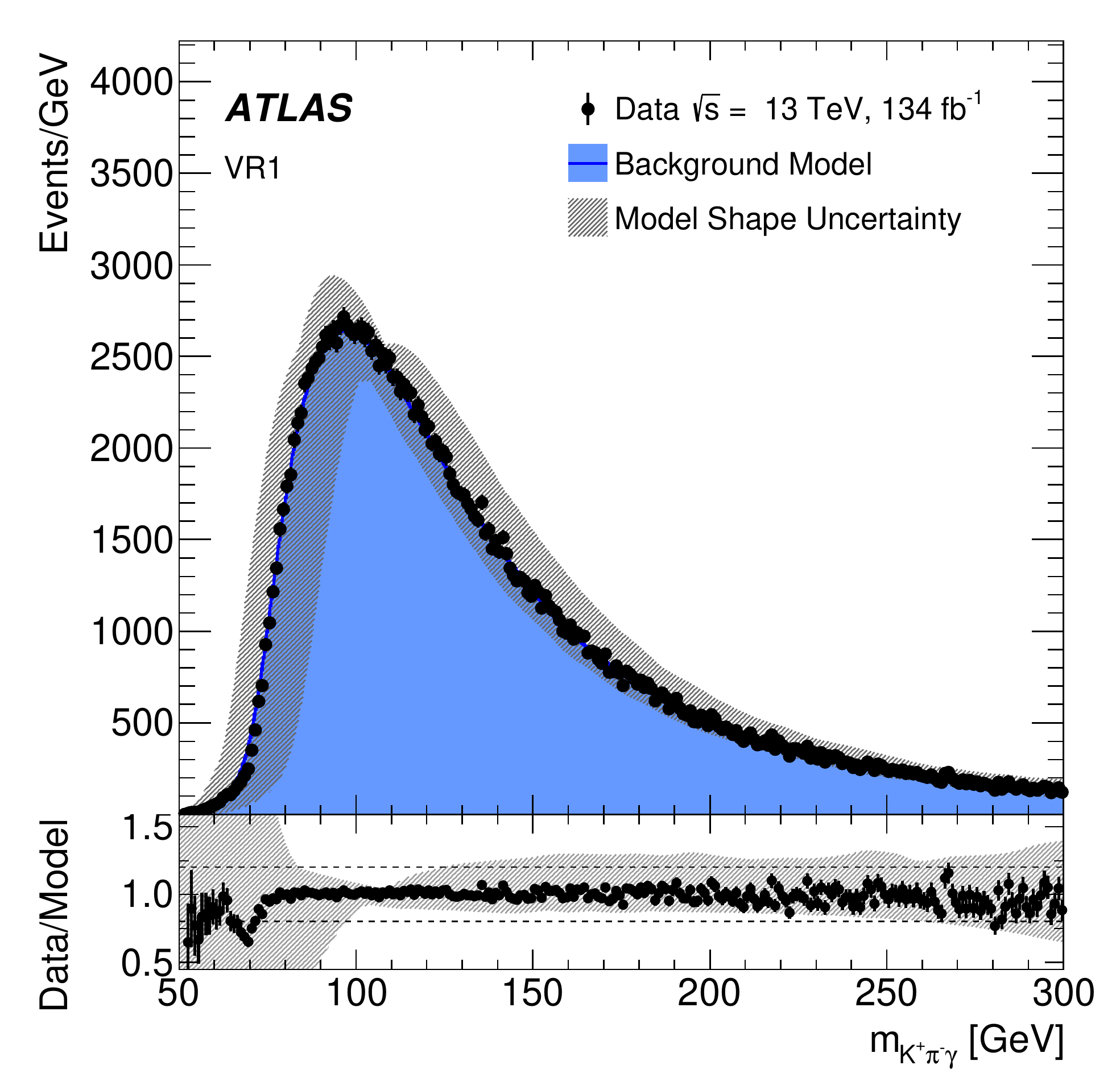}\label{fig:KStarvr1}}
\subfigure[]{\includegraphics[width=0.35\textwidth]{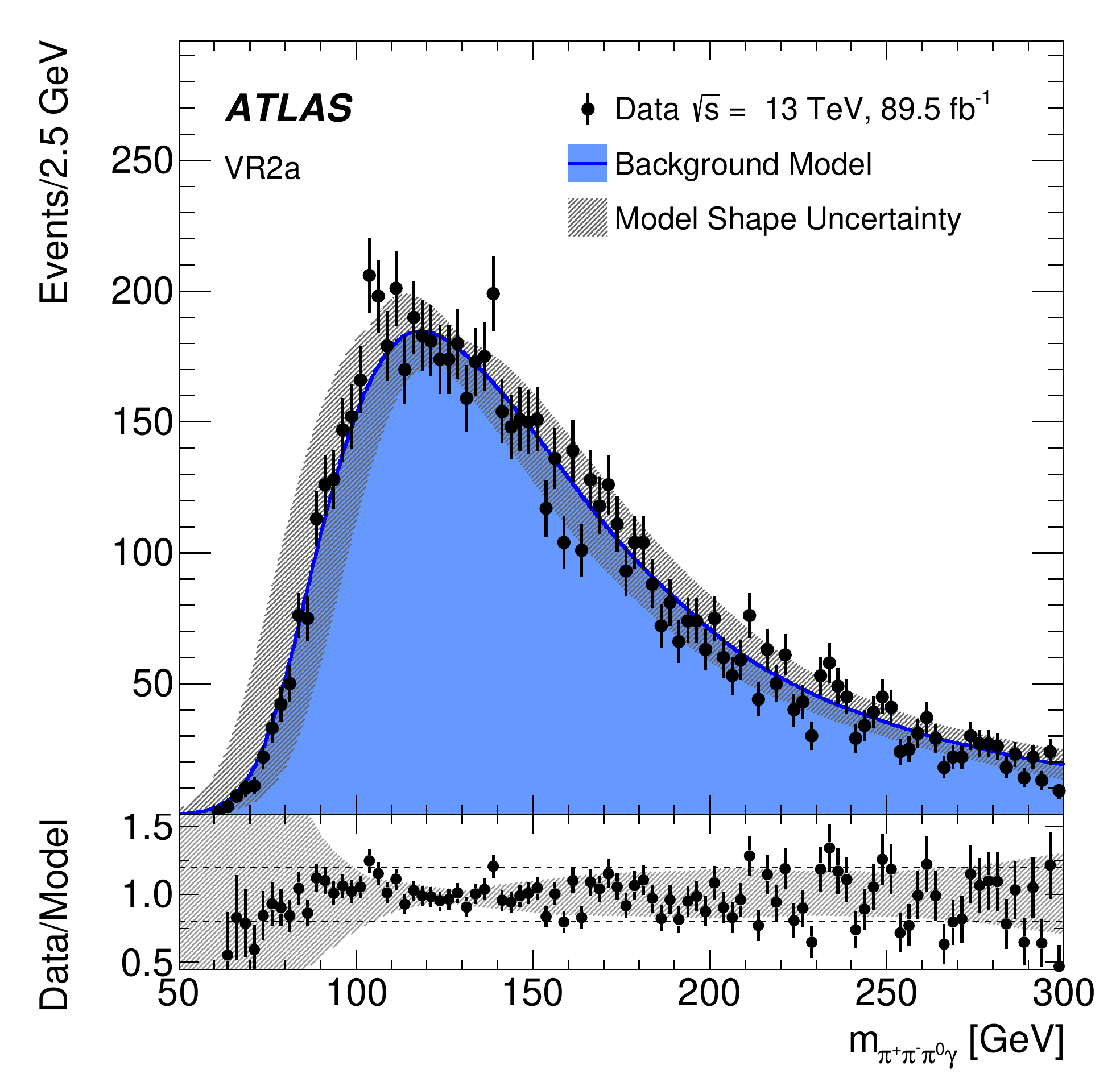}\label{fig:Omegavr2}}
\subfigure[]{\includegraphics[width=0.35\textwidth]{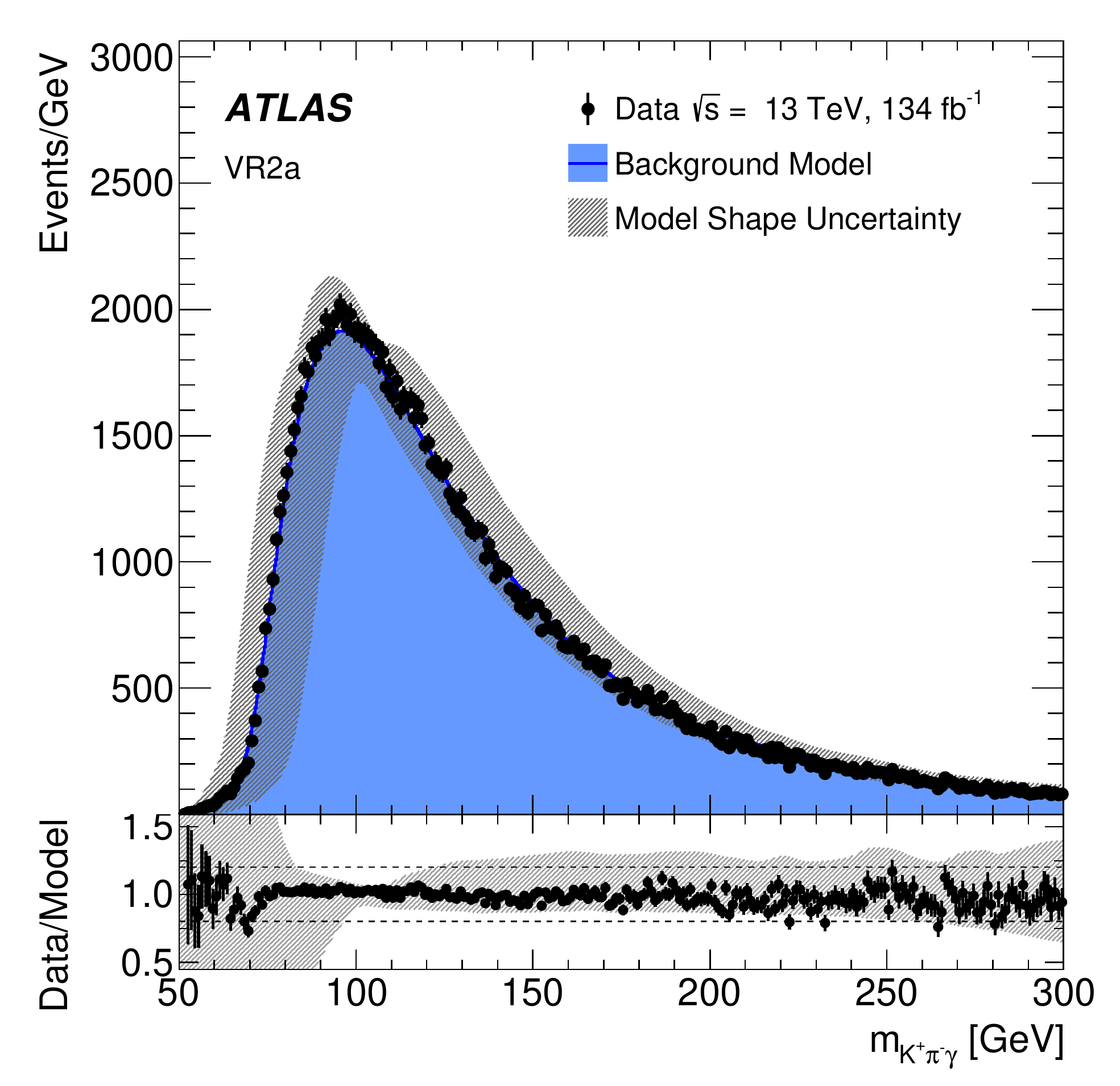}\label{fig:KStarvr2}}
\subfigure[]{\includegraphics[width=0.35\textwidth]{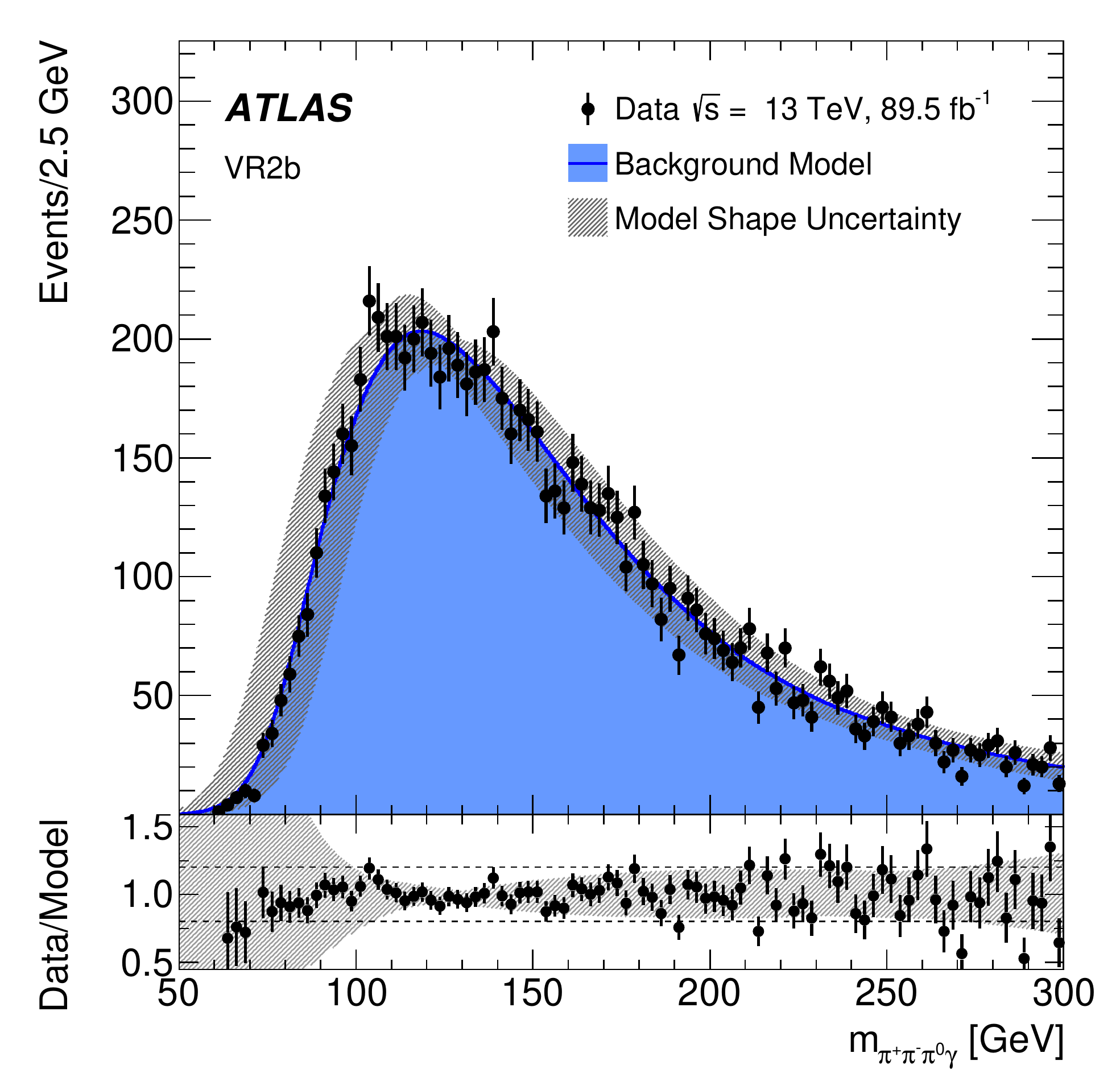}\label{fig:Omegavr3}}
\subfigure[]{\includegraphics[width=0.35\textwidth]{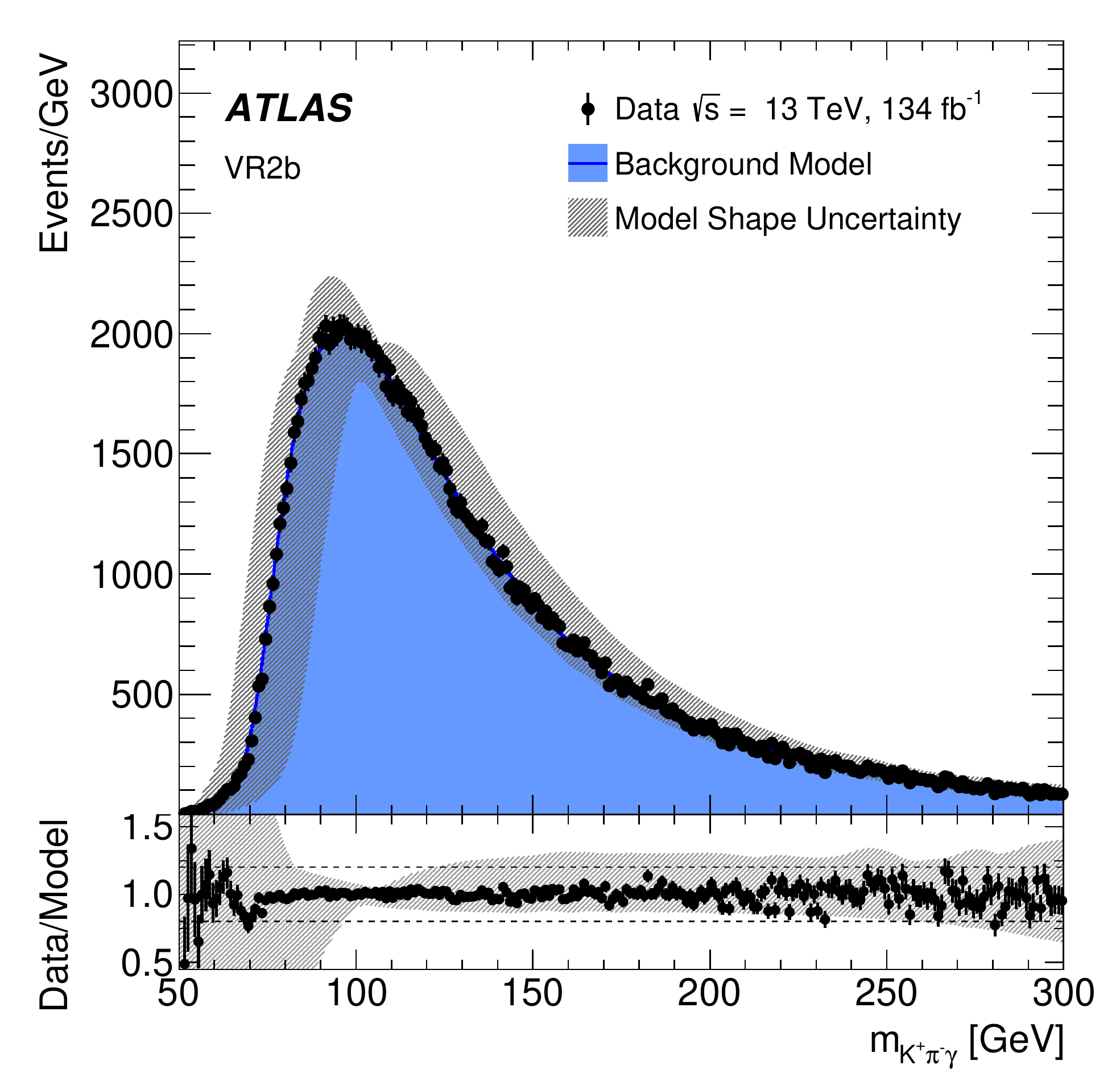}\label{fig:KStarvr3}}
\caption{Distributions of (\subref{fig:Omegavr1}, \subref{fig:Omegavr2}, \subref{fig:Omegavr3}) $m_{\pi\pi\pi\gamma}$ and (\subref{fig:KStarvr1}, \subref{fig:KStarvr2}, \subref{fig:KStarvr3}) $m_{K\pi\gamma}$ in data compared with the prediction of the background
model for the VR1, VR2a and VR2b validation regions. The background model is normalised to the observed number
of events within the region shown. The uncertainty band corresponds to the uncertainty envelope derived from
variations in the background modelling procedure, described in Section~\ref{sec:background}. The ratio of the data to the background model is shown below
the distributions.}
\label{fig:BkgdModelValidation}
\end{figure}

To validate the background model, the $m_{\mathcal{M}\gamma}$ distributions in validation regions, defined by kinematic and isolation requirements looser than the nominal signal requirements, are used to compare the prediction of the background model with the data. Three validation regions are defined, each based on the GR selection and adding one of the following: the meson isolation requirement (VR1), the calorimeter component of the photon isolation requirements (VR2a), or the track component of the photon isolation requirements (VR2b). The $m_{\mathcal{M}\gamma}$ distributions in these validation regions are shown in Figure~\ref{fig:BkgdModelValidation}. The background model is found to describe the data well in all regions within uncertainties.
 
To allow the shape of the background model for the three-body mass to adjust to the observed data, variations around the nominal
shape are derived. Three variations are included to allow this adjustment to occur. Firstly, a scale variation of the $\pt$ distribution
of either $\mathcal{M}$ or $\gamma$ is used to allow the peak of the three-body mass distribution to shift upwards or downwards in mass.
Secondly, a variation is produced via a polynomial distortion of the shape of the $\Delta\phi(\mathcal{M},\gamma)$ distribution, which allows
the width of the three-body mass distribution to vary. Finally, a global tilt of the three-body mass distribution is included, which allows
the model to account for slopes with respect to the data. The first two variations are simple alterations to the kinematics of the
pseudocandidates, resulting in changes to the three-body mass distribution. The corresponding nuisance parameters are assigned a Gaussian constraint
in the likelihood. The third variation is directly applied to the three-body mass template and is not constrained by a Gaussian term in the fit.
The uncertainty band in Figure~\ref{fig:BkgdModelValidation} corresponds to the uncertainty envelope derived from
these variations, which are set sufficiently large to allow the shape to adapt in the fit to the data.
 
\section{Systematic uncertainties}
\label{sec:systematics}
 
The photon identification and isolation uncertainties are estimated to be 1.7$\%$ for the Higgs and $Z$ boson signals. An uncertainty of 3.0$\%$ per $\mathcal{M}$ candidate is assigned to the track reconstruction efficiency and accounts for effects associated with the modelling of ID material and also with the track reconstruction algorithms if there is a charged particle near the photon. This uncertainty is derived conservatively by assuming the uncertainty to be fully correlated between the two tracks of the $\mathcal{M}$ candidate.
 
The systematic uncertainty in the expected signal yield arising from
the photon component of the trigger efficiency is estimated to be 0.7$\%$~\cite{TRIG-2018-05}. Uncertainties to
account for the \enquote{$\tau$}-component of the trigger turn on are assigned
in accord with the studies described in Ref.~\cite{JETM-2018-08}, and are
7.3$\%$ (2.1$\%$) for the $H(Z)\to\omega\gamma$ channels, and 4.1$\%$
for the $H\to K^{*}\gamma$ channel.
 
The Higgs boson production cross sections and decay branching fractions, as
well as their uncertainties, are taken from
Refs.~\cite{LHCHiggsCrossSectionWorkingGroup:2011ti,LHCHiggsCrossSectionWorkingGroup:2012vm,Heinemeyer:2013tqa}. The effect of QCD scale uncertainties on the cross section for a 125~\GeV\ $H$ boson~\cite{Heinemeyer:2013tqa} amounts to a 5.0$\%$ uncertainty. The uncertainties in the
production cross section due to uncertainties in the PDFs and the strong coupling constant, \alphas, are combined for the separate processes and amount to $2.9\%$. The shape component of the uncertainties was found to be negligible.
The efficiency to reconstruct the neutral pion in the decay using the TauPFO algorithms is approximately 65$\%$. The uncertainty in this reconstruction efficiency, based on Ref.~\cite{PERF-2017-02}, is 5.0$\%$.
 
The shape of the background model is allowed to vary around the nominal shape, and the parameters controlling these systematic variations are treated as nuisance parameters in the maximum-likelihood fit used to extract the signal and background yields, as described in Section~\ref{sec:BkgdValidation}.

\section{Results}
\label{sec:result}
After the full selection is applied, 4264 events are observed in the $H/Z\to\omega\gamma$ signal region, and 114\,707 events are observed in the $H\to K^{*}\gamma$ signal region. The data are compared with background and signal predictions using an unbinned maximum-likelihood fit to the $m_{\mathcal{M}\gamma}$ distribution. The parameters of interest are the Higgs and $Z$ boson signal normalisations. Systematic uncertainties are modelled using additional nuisance parameters in the fit; in particular, the background normalisation is a free parameter in the model. The fit uses the selected events with $m_{\mathcal{M}\gamma} < 300$~\GeV.
Upper limits are set on the branching fractions for the Higgs and $Z$ boson decays into $\mathcal{M}\gamma$ using the CL$_\text{s}$ modified frequentist formalism~\cite{Read:2002hq} with the profile-likelihood-ratio test statistic and the asymptotic approximations derived in Ref.~\cite{Cowan:2010js}. For the upper limits on the branching fractions, the SM production cross section is assumed for the Higgs boson, while the ATLAS measurement of the inclusive $Z$ boson cross section is used for the $Z$ boson signal, as discussed in Section~\ref{sec:systematics}. Tests of the fit on sideband meson-mass data demonstrate good agreement and modelling of the background processes and serve as further validation of the background modelling.
 
\begin{figure}[!htbp]
\centering
\subfigure[]{\includegraphics[width=0.49\textwidth]{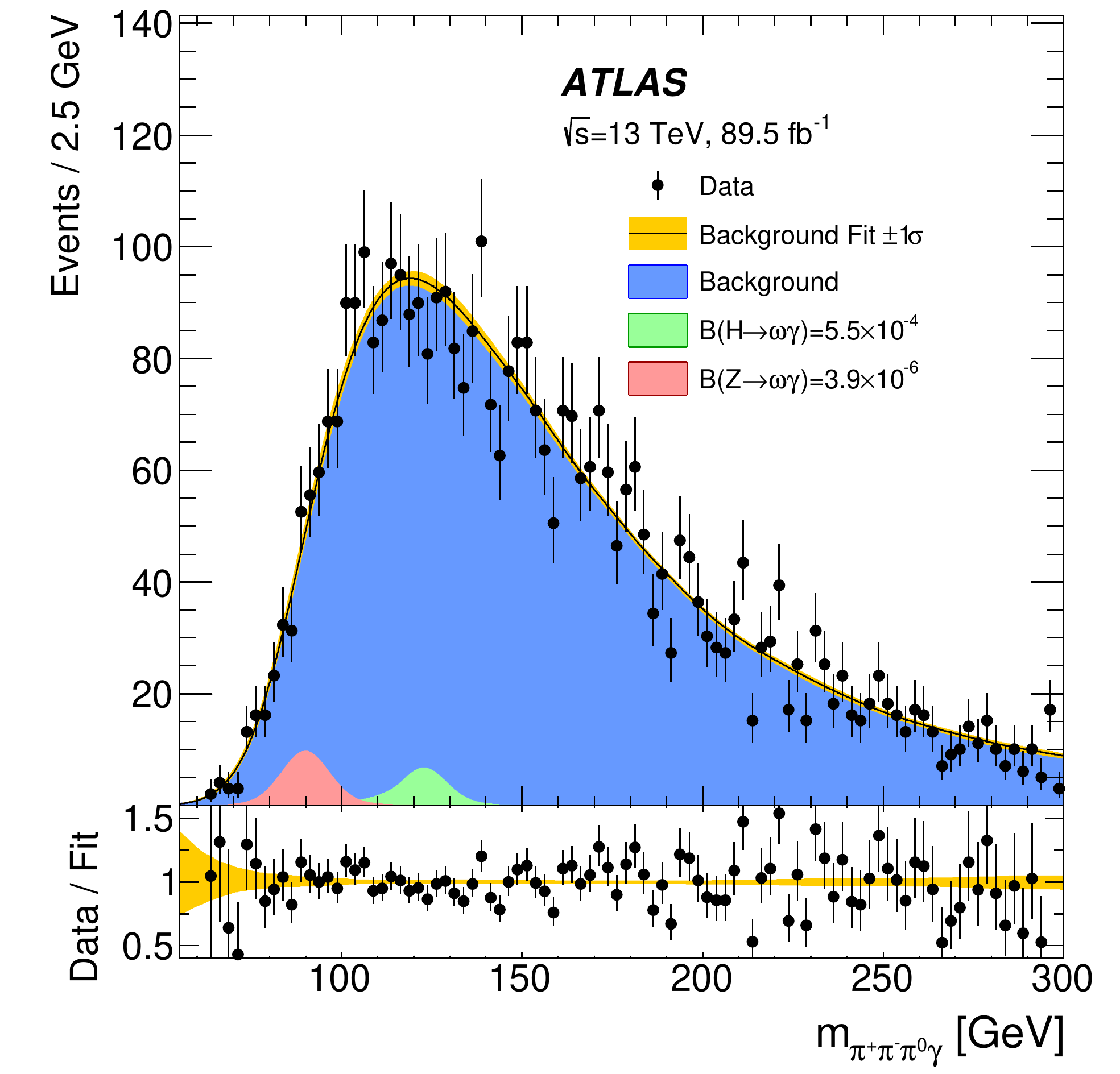}\label{fig:Omega}}
\subfigure[]{\includegraphics[width=0.49\textwidth]{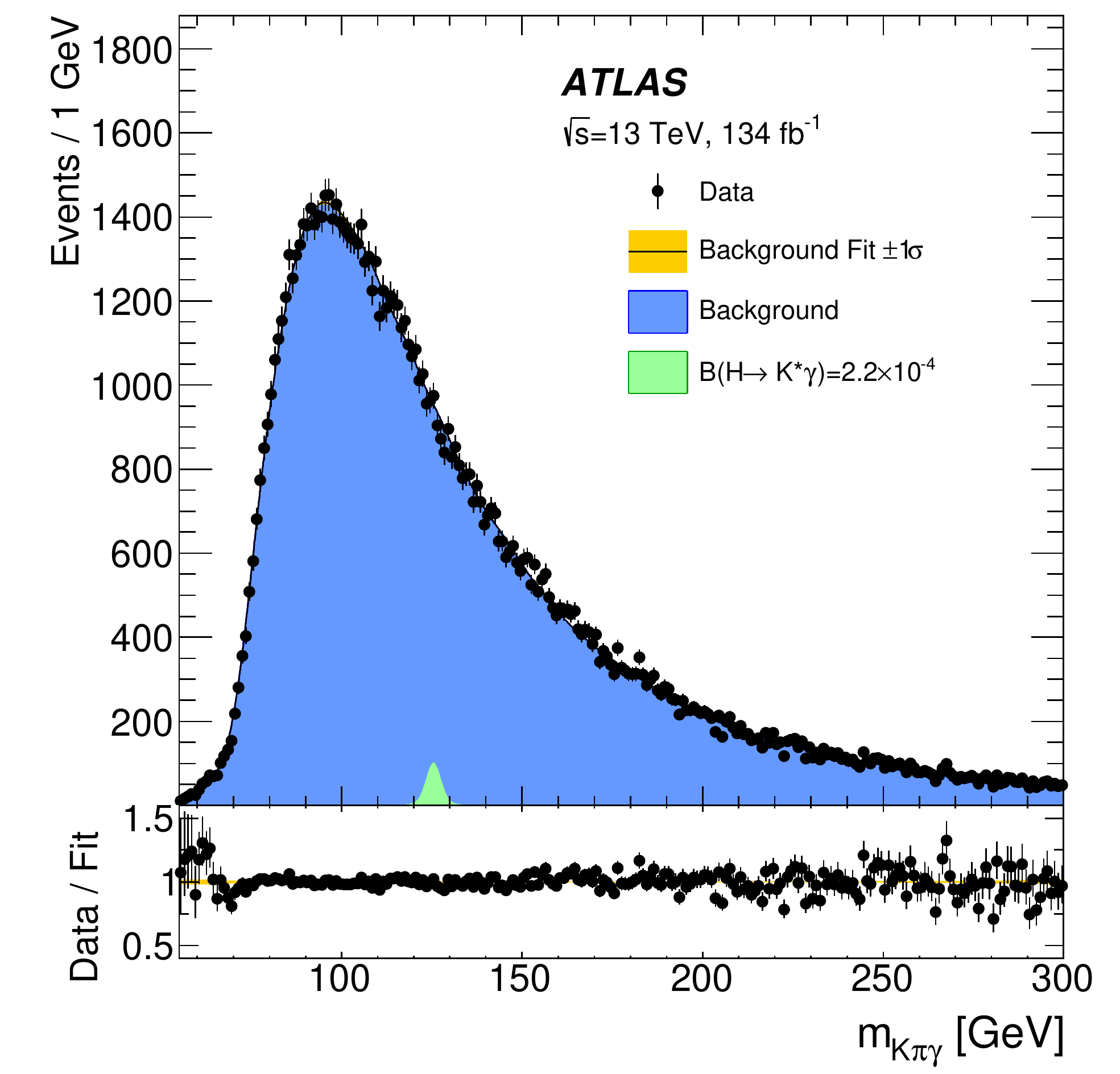}\label{fig:KStar}}
\caption{Background-only fits performed in the signal region for the \subref{fig:Omega} $\omega\gamma$ and \subref{fig:KStar} $K^{*}\gamma$ final states.
The branching fraction of each of the signals is set to the observed $95\%$ CL upper limit.
The yellow band represents the uncertainty in the fit arising from the constrained background shape systematic uncertainties.}
\label{fig:Fits}
\end{figure}

The expected and observed numbers of background events within regions of $m_{\mathcal{M}\gamma}$ around the Higgs and $Z$ boson masses are given in Table~\ref{tab:events}. The expected numbers of background events are taken from background-only fits. Table~\ref{tab:events} also shows the expected number of signal events for branching fractions near the sensitivity of the analyses.
Fits to the signal region data are shown in Figure~\ref{fig:Fits}. Signals are shown normalised to branching fraction values of the order of the expected sensitivity. The observed and expected limits are summarised in Table~\ref{tab:OmegaGammaLimits}.
\begin{table}[!htbp]
\centering
\caption{Numbers of observed and expected background events for the $m_{\mathcal{M}\gamma}$ ranges of interest.
Each expected background and the corresponding uncertainty of its mean is obtained from a background-only fit to the data; the uncertainty does not take into account statistical fluctuations in each mass range.
Expected $Z$ and Higgs boson signal contributions, with their corresponding total systematic uncertainty, are shown for reference branching fractions of $10^{-6}$ and $10^{-4}$, respectively.
Entries are marked with a dash when there is no signal of that type in the specified range.}\label{tab:events}
\begin{tabular}{lcccc}
\hline
Channel            & Mass range  & Observed (Expected)  & $H$ signal & $Z$ signal  \\
& [\GeV] & background &  ${\cal B}=10^{-4}$ & ${\cal B}=10^{-6}$ \\ \hline
$H\to\omega\gamma$ & 115--135  & \phantom{0}686 \phantom{0}$(730 \pm 17)$     & 9 $\pm$ 1    & --              \\
$Z\to\omega\gamma$ & \phantom{1}80--100   & \phantom{0}388 \phantom{0}$(386 \pm 16)$     & --               & 18 $\pm$ 2    \\
$H\to K^{*}\gamma$ & 120--130& 9526 $(9630 \pm 50)$  & 53 $\pm$ 4  & --\\  \hline
\end{tabular}
\end{table}
 
\begin{table}[!htbp]
\centering
\caption{Expected and observed branching fraction limits at the 95$\%$ CL for $H/Z\rightarrow \omega\gamma$ and $H\rightarrow K^{*}\gamma$.}
\label{tab:OmegaGammaLimits}
\begin{tabular}{lcc}
\hline
\multicolumn{1}{l}{Channel} & \multicolumn{2}{c}{95\% CL upper limit}                    \\
& Expected               & Observed             \\ \hline
$H\to\omega\gamma$ [$10^{-4}$]   & $\tabscript{10.4}{+3.8}{-2.9}$ & $5.5$ \\
$Z\to\omega\gamma$ [$10^{-6}$]   & \phantom{0}$\tabscript{4.7}{+2.0}{-1.3}$ & $3.9$ \\
$H\to K^{*}\gamma$ [$10^{-4}$]   & \phantom{0}$\tabscript{3.7}{+1.5}{-1.0}$ & $2.2$ \\\hline
\end{tabular}
\end{table}

\section{Conclusion}
\label{sec:conclusion}
 
A search for the decays $H/Z\to\omega\gamma$ and $H\to K^{*}\gamma$ has been performed with 13~\TeV\
$pp$ collision data samples collected with the ATLAS detector at the LHC corresponding to integrated luminosities of 89.5~fb$^{-1}$ and 134~fb$^{-1}$ respectively. The $\omega$ and $K^{*}$ mesons are reconstructed via their dominant decays into $\pi^{+}\pi^{-}\pi^{0}$ and $K^{\pm}\pi^{\pm}$ final states, respectively. The background model is derived using a fully data-driven approach and validated in a number of different regions.
No significant excess of events above the SM background expectations is observed. The obtained 95$\%$ CL upper limits are ${\mathcal{B}}(H\to\omega\gamma) < 5.5\times 10^{-4}$ (370$\times$SM), ${\mathcal{B}}(Z\to\omega\gamma) < 3.9\times 10^{-6}$ (140$\times$SM), and ${\mathcal{B}}(H\to K^{*}\gamma) < 2.2\times 10^{-4}$. The result for $Z\rightarrow \omega\gamma$ corresponds to a two-orders-of-magnitude improvement over the previously set limit at DELPHI.
 
\section*{Acknowledgements}


We thank CERN for the very successful operation of the LHC, as well as the
support staff from our institutions without whom ATLAS could not be
operated efficiently.
 
We acknowledge the support of
ANPCyT, Argentina;
YerPhI, Armenia;
ARC, Australia;
BMWFW and FWF, Austria;
ANAS, Azerbaijan;
CNPq and FAPESP, Brazil;
NSERC, NRC and CFI, Canada;
CERN;
ANID, Chile;
CAS, MOST and NSFC, China;
Minciencias, Colombia;
MEYS CR, Czech Republic;
DNRF and DNSRC, Denmark;
IN2P3-CNRS and CEA-DRF/IRFU, France;
SRNSFG, Georgia;
BMBF, HGF and MPG, Germany;
GSRI, Greece;
RGC and Hong Kong SAR, China;
ISF and Benoziyo Center, Israel;
INFN, Italy;
MEXT and JSPS, Japan;
CNRST, Morocco;
NWO, Netherlands;
RCN, Norway;
MEiN, Poland;
FCT, Portugal;
MNE/IFA, Romania;
MESTD, Serbia;
MSSR, Slovakia;
ARRS and MIZ\v{S}, Slovenia;
DSI/NRF, South Africa;
MICINN, Spain;
SRC and Wallenberg Foundation, Sweden;
SERI, SNSF and Cantons of Bern and Geneva, Switzerland;
MOST, Taiwan;
TENMAK, T\"urkiye;
STFC, United Kingdom;
DOE and NSF, United States of America.
In addition, individual groups and members have received support from
BCKDF, CANARIE, Compute Canada and CRC, Canada;
PRIMUS 21/SCI/017 and UNCE SCI/013, Czech Republic;
COST, ERC, ERDF, Horizon 2020, ICSC-NextGenerationEU and Marie Sk{\l}odowska-Curie Actions, European Union;
Investissements d'Avenir Labex, Investissements d'Avenir Idex and ANR, France;
DFG and AvH Foundation, Germany;
Herakleitos, Thales and Aristeia programmes co-financed by EU-ESF and the Greek NSRF, Greece;
BSF-NSF and MINERVA, Israel;
Norwegian Financial Mechanism 2014-2021, Norway;
NCN and NAWA, Poland;
La Caixa Banking Foundation, CERCA Programme Generalitat de Catalunya and PROMETEO and GenT Programmes Generalitat Valenciana, Spain;
G\"{o}ran Gustafssons Stiftelse, Sweden;
The Royal Society and Leverhulme Trust, United Kingdom.
 
The crucial computing support from all WLCG partners is acknowledged gratefully, in particular from CERN, the ATLAS Tier-1 facilities at TRIUMF (Canada), NDGF (Denmark, Norway, Sweden), CC-IN2P3 (France), KIT/GridKA (Germany), INFN-CNAF (Italy), NL-T1 (Netherlands), PIC (Spain), ASGC (Taiwan), RAL (UK) and BNL (USA), the Tier-2 facilities worldwide and large non-WLCG resource providers. Major contributors of computing resources are listed in Ref.~\cite{ATL-SOFT-PUB-2023-001}.


\printbibliography
 
\clearpage
 
\clearpage
\input{atlas_authlist}

\end{document}

%% file: atlas_authlist.tex
 
\begin{flushleft}
\hypersetup{urlcolor=black}
{\Large The ATLAS Collaboration}

\bigskip

\AtlasOrcid[0000-0002-6665-4934]{G.~Aad}$^\textrm{\scriptsize 102}$,
\AtlasOrcid[0000-0002-5888-2734]{B.~Abbott}$^\textrm{\scriptsize 120}$,
\AtlasOrcid[0000-0002-1002-1652]{K.~Abeling}$^\textrm{\scriptsize 55}$,
\AtlasOrcid[0000-0002-8496-9294]{S.H.~Abidi}$^\textrm{\scriptsize 29}$,
\AtlasOrcid[0000-0002-9987-2292]{A.~Aboulhorma}$^\textrm{\scriptsize 35e}$,
\AtlasOrcid[0000-0001-5329-6640]{H.~Abramowicz}$^\textrm{\scriptsize 151}$,
\AtlasOrcid[0000-0002-1599-2896]{H.~Abreu}$^\textrm{\scriptsize 150}$,
\AtlasOrcid[0000-0003-0403-3697]{Y.~Abulaiti}$^\textrm{\scriptsize 117}$,
\AtlasOrcid[0000-0003-0762-7204]{A.C.~Abusleme~Hoffman}$^\textrm{\scriptsize 137a}$,
\AtlasOrcid[0000-0002-8588-9157]{B.S.~Acharya}$^\textrm{\scriptsize 69a,69b,p}$
% ...
\bigskip
\\

% ...
$^{29}$Physics Department, Brookhaven National Laboratory, Upton NY; United States of America.\\
% ...
$^{35}$$^{(a)}$Facult\'e des Sciences Ain Chock, R\'eseau Universitaire de Physique des Hautes Energies - Universit\'e Hassan II, Casablanca;$^{(b)}$Facult\'{e} des Sciences, Universit\'{e} Ibn-Tofail, K\'{e}nitra;$^{(c)}$Facult\'e des Sciences Semlalia, Universit\'e Cadi Ayyad, LPHEA-Marrakech;$^{(d)}$LPMR, Facult\'e des Sciences, Universit\'e Mohamed Premier, Oujda;$^{(e)}$Facult\'e des sciences, Universit\'e Mohammed V, Rabat;$^{(f)}$Institute of Applied Physics, Mohammed VI Polytechnic University, Ben Guerir; Morocco.\\
% ...
$^{55}$II. Physikalisches Institut, Georg-August-Universit\"{a}t G\"ottingen, G\"ottingen; Germany.\\
% ...
$^{69}$$^{(a)}$INFN Gruppo Collegato di Udine, Sezione di Trieste, Udine;$^{(b)}$ICTP, Trieste;$^{(c)}$Dipartimento Politecnico di Ingegneria e Architettura, Universit\`a di Udine, Udine; Italy.\\
% ...
$^{102}$CPPM, Aix-Marseille Universit\'e, CNRS/IN2P3, Marseille; France.\\
% ...
$^{117}$Department of Physics, New York University, New York NY; United States of America.\\
% ...
$^{120}$Homer L. Dodge Department of Physics and Astronomy, University of Oklahoma, Norman OK; United States of America.\\
% ...
$^{137}$$^{(a)}$Departamento de F\'isica, Pontificia Universidad Cat\'olica de Chile, Santiago;$^{(b)}$Millennium Institute for Subatomic physics at high energy frontier (SAPHIR), Santiago;$^{(c)}$Instituto de Investigaci\'on Multidisciplinario en Ciencia y Tecnolog\'ia, y Departamento de F\'isica, Universidad de La Serena;$^{(d)}$Universidad Andres Bello, Department of Physics, Santiago;$^{(e)}$Instituto de Alta Investigaci\'on, Universidad de Tarapac\'a, Arica;$^{(f)}$Departamento de F\'isica, Universidad T\'ecnica Federico Santa Mar\'ia, Valpara\'iso; Chile.\\
% ...
$^{150}$Department of Physics, Technion, Israel Institute of Technology, Haifa; Israel.\\
$^{151}$Raymond and Beverly Sackler School of Physics and Astronomy, Tel Aviv University, Tel Aviv; Israel.\\
% ...
$^{a}$ Also Affiliated with an institute covered by a cooperation agreement with CERN.\\
$^{b}$ Also at An-Najah National University, Nablus; Palestine.\\
$^{c}$ Also at Borough of Manhattan Community College, City University of New York, New York NY; United States of America.\\
$^{d}$ Also at Bruno Kessler Foundation, Trento; Italy.\\
$^{e}$ Also at Center for High Energy Physics, Peking University; China.\\
$^{f}$ Also at Center for Interdisciplinary Research and Innovation (CIRI-AUTH), Thessaloniki; Greece.\\
% ...
$^{p}$ Also at Department of Physics, King's College London, London; United Kingdom.\\
% ...

\end{flushleft}
